\documentclass[aps,prl,twocolumn,groupedaddress,amsmath,nofootinbib]{revtex4-1}

\usepackage{amssymb}
\usepackage{amsfonts}
\usepackage{amssymb}
\usepackage{amsmath}
\usepackage[dvips]{graphicx}
\usepackage{color}

\usepackage{multirow}

\usepackage{txfonts}
\usepackage{hyperref}

\begin{document}

\title{Quantum beam splitter for orbital angular momentum of light: quantum correlation by four-wave mixing operated in a nonamplifying regime}

\author{Wei Liu$^{1}$}
\author{Rong Ma$^{1}$}
\author{Li Zeng$^{1}$}
\author{Zhongzhong Qin$^{1,2}$}
\email{zzqin@sxu.edu.cn}
\author{Xiaolong Su$^{1,2}$}
\email{suxl@sxu.edu.cn}

\affiliation{$^1$State Key Laboratory of Quantum Optics and Quantum Optics Devices,
Institute of Opto-Electronics, Shanxi University, Taiyuan 030006, People's
Republic of China\\
$^2$Collaborative Innovation Center of Extreme Optics, Shanxi University,
Taiyuan, Shanxi 030006, People's Republic of China}
\date{\today}

\begin{abstract}
Nondegenerate four-wave mixing (FWM) process based on a double-$\Lambda$ scheme in hot alkali metal vapor is a versatile tool in quantum state engineering, quantum imaging, and quantum precision measurements. In this Letter, we investigate the generation of quantum correlated twin beams which carry nonzero orbital angular momentums (OAMs) based on the FWM process in hot cesium vapor. The amplified probe beam and the newly generated conjugate beam in the FWM process have the same and opposite topological charge as the seed beam, respectively. We also explore the FWM process operated in a nonamplifying regime where quantum correlated twin beams carrying OAMs can still be generated. In this regime, the FWM process plays the role of quantum beam splitter for the OAM of light, that is, a device that can split a coherent light beam carrying OAM into quantum-correlated twin beams carrying OAMs. More generally, our setup can be used as a quantum beam splitter of images.
\end{abstract}

\maketitle

Quantum correlation and entanglement attract increasing interest due to their significance for fundamental tests of quantum physics \cite{EPR,Bell}, and potential applications in
future quantum technologies \cite{GravitationalWave,QuantumImaging,BraunsteinRMP}. This is especially true in optics because light is an ideal carrier of information. 
The most commonly used method to generate squeezed states and continuous-variable Einstein-Podolsky-Rosen (EPR) entangled states is by parametric down-conversion in a nonlinear crystal, with an optical parametric oscillator (OPO) or optical parametric amplifier \cite{SchnabelPRL,JiaOE}.

On the other hand, the first generation of squeezed state was realized by four-wave mixing (FWM) process in sodium vapor \cite{SlusherPRL}. However, the maximal degree of squeezing generated from atomic ensemble was no more than 2.2 dB in the following twenty years, mainly due to the limitation of spontaneous emission and absorption in atomic vapor \cite{AtomsSqueezing7}. Unitl 2007, it has been shown by several groups that the FWM process based on a double-$\Lambda$ scheme in hot rubidium vapor is an efficient way to generate quantum correlated twin beams with strong quantum noise reduction \cite{LettOL,LettPRA,QinOL,QuentinPRA,JasperseOE}. A maximum of 9.2 dB intensity-difference squeezing has been realized in this system \cite{QuentinPRA}. The substantial noise reduction benefits from the coherence effects between the hyperfine electronic ground states, low spontaneous emission rate thanks to a relatively small atomic population in the excited states, as well as a judicious choice of phase-matching condition \cite{PhaseMatching}. The central frequency and linewidth of the generated squeezed states from the FWM process intrinsically match atomic transitions. In addition, the lack of a cavity makes the system immune to environmental noise, and allows the system to operate as a multi-spatial-mode phase-insensitive amplifier. These advantages make this system very successful for a variety of applications such as the generation of multiple quantum correlated beams \cite{QinPRL,QinAPL}, entangled images \cite{EntangledImages}, high purity single photons \cite{QinLight}, as well as optical qubits \cite{TravisOL}, the tunable delay of Einstein-Podolsky-Rosen (EPR) entangled states \cite{TunableDelay}, the realization of a SU(1,1) nonlinear interferometer \cite{SU11NC}, and the ultrasensitive measurement of microcantilever displacement below the shot-noise limit \cite{PooserDisplacement}.

It has been shown that quantum correlated twin beams with 6.5 dB noise reduction can also be generated from FWM process in hot cesium vapor by our group \cite{MaOL,MaPRA}. Cesium offers certain advantages, e.g., the quantum correlation at the $^{133}$Cs $D_{1}$ line lies well within the wavelength regime of the exciton emission from InAs quantum dots \cite{QuantumDots}, which provides a potential resource for coherent interfaces between atomic and solid-state systems.

In the past decades, orbital angular momentum (OAM) of light attracts increasing interest because of its great potential in enhancing the information channel capacity in both classical and quantum optical communications \cite{OAMReview1,OAMReview2,DSDingPRL}. Continuous variable quadrature entanglement between two $l=\pm1$ order Laguerre-Gaussian (LG) modes has been generated using a spatially nondegenerate OPO \cite{AndersenPRL}. The generation of quantum correlated twin beams carrying $l=\pm1$ order OAM respectively, as well as $l=0$ and $l=+2$ order OAM respectively has been demonstrated based on the FWM process in hot rubidium vapor with different configurations of OAM \cite{LettOAM, LettOAM2}.

\begin{figure}[t]
\centering
\fbox{\includegraphics[width=\linewidth]{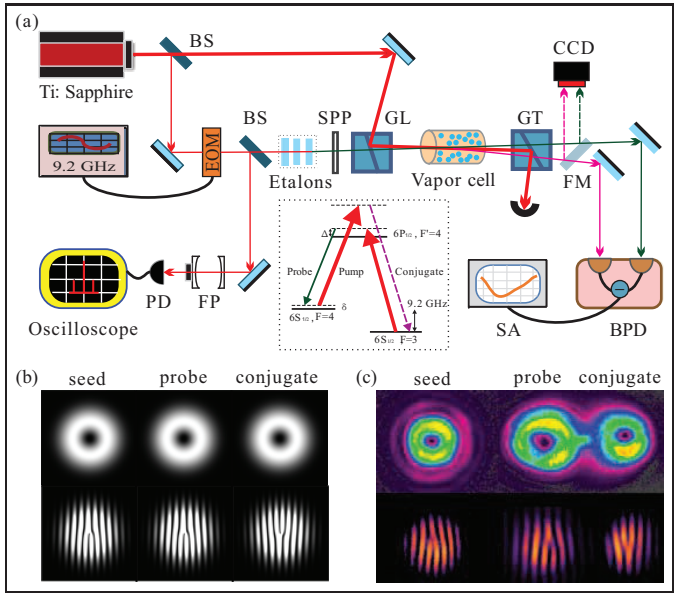}}
\caption{Experimental schematic for generating and detecting quantum correlated twin beams carrying OAMs. (a) Experimental setup. BS: beam splitter; EOM: electro-optic modulator; FP: Fabry-Perot interferometer; PD: photodetector; SPP: spiral phase plate; GL: Glan-laser polarizer; GT: Glan-Thompson polarizer; FM: flip mirror; CCD: charge-coupled device camera; BPD: balanced photodetector; SA: spectrum analyzer. Inset: Double-$\Lambda$ scheme in the $D_{1}$ line of $^{133}$Cs. $\Delta$ is one-photon detuning and $\delta$ is two-photon detuning. (b) Upper panel: Ideal beam patterns of the seed, probe, and conjugate beams. Lower panel: Corresponding interference patterns with a plane wave beam. (c) Upper panel: Experimentally measured beam patterns of the seed, probe, and conjugate beams. Lower panel: Experimentally measured interference patterns with a plane wave beam.}
\end{figure}

In this Letter, we first experimentally demonstrated the generation of quantum correlated twin beams which carry nonzero OAMs based on the double-$\Lambda$ scheme FWM process in hot cesium vapor. In particular, the seed beam is a LG beam which carries topological charge $l=-1$. After the FWM process, the amplified probe beam has a topological charge $l=-1$, while the newly generated conjugate beam has a topological charge $l=+1$, so that the conservation of angular momentum is satisfied in the FWM process. In Ref. \cite{QuentinPRA}, the authors demonstrated that the twin beams generated from the FWM process operated in a nonamplifying regime are also quantum correlated, which they named as "quantum beam splitter for photons". Motivated by Ref. \cite{QuentinPRA}, we explored the case in which a LG coherent beam carrying topological charge $l=-1$ is seeded into the FWM process in hot cesium vapor operated in a nonamplifying regime. Interestingly, quantum correlated twin beams carrying OAMs are also generated when the total gain of the probe and conjugate beams is one. The probe beam and conjugate beam after the FWM process are in analogy to two output channels of an optical beam splitter. However, different from an optical beam splitter, these twin beams are quantum correlated and carry $l=-1$ and $l=+1$ OAM, respectively. In this sense, our setup can be used as a quantum beam splitter for OAM of light. The effective transmission and reflection coefficients of our quantum beam splitter can be manipulated within a certain range by tuning the two-photon detuning of the FWM process.

As shown in the inset of Fig. 1(a), the $^{133}$Cs $D_{1}$ line is used to form the double-$\Lambda$ level structure with an excited level
($6P_{1/2}, F'=4$) and two ground levels ($6S_{1/2}, F=3$ and $F=4$). In the FWM process, two pump photons are simultaneously converted to one probe photon and one conjugate photon. As a result, the relative intensity difference of the probe and conjugate beams is squeezed compared with the corresponding shot-noise limit (SNL) by an amount of $1/(2G-1)$, where $G$ is the gain of the FWM process. The Ti:sapphire laser is tuned about 1.6 GHz to the blue of $^{133}$Cs ($6S_{1/2}, F=3\rightarrow6P_{1/2}, F'=4$) with a total power of 1 W. The laser beam is split into two beams by a beam splitter (BS). One of the beams serves as the pump beam with a power of around 550 mW, and the other beam passes through an electro-optic modulator (EOM) to produce optical sidebands at $\pm$9.2 GHz from the carrier frequency. The modulation efficiency of the EOM is monitored by a scanning Fabry-Perot (FP) interferometer. Three successive temperature-stabilized etalons (with a thickness of 7 mm, 7 mm, and 3 mm, respectively)  are used to select the probe frequency component ($-1^{st}$ order sideband), which provide a combined extinction ratio of over 40 dB \cite{MaOL}. Then the beam passes through a spiral phase plate (SPP) to generate a coherent beam carrying topological charge $l=-1$.

\begin{figure}[t]
\centering
\fbox{\includegraphics[width=\linewidth]{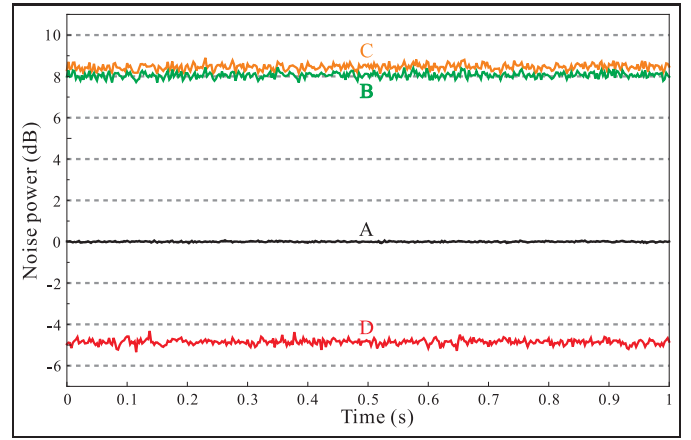}}
\caption{Normalized noise power of the probe beam (trace $B$), conjugate beam (trace $C$), and the intensity-difference of the twin beams (trace $D$). Trace $A$ at 0 dB shows the corresponding SNL of trace $B\sim D$. Experimental parameters are as follows: $\Delta = 1.6$ GHz,  $\delta = 0$ MHz, $P_{Pump}$ = 550 mW, and $T$ = 112 $^{\circ}$C. Resolution bandwidth (RBW) of SA is 30 kHz and video bandwidth (VBW) of SA is 100 Hz.}
\end{figure}

The seed beam and the pump beam cross each other in the center of the cesium vapor cell at an angle of 6 mrad. The vapor cell is 25 mm long and its temperature is stabilized at $112 ^{\circ}$C. In order to ensure the  pump beam and the probe beam overlap over almost the full length of the cell, their waists are 780 and 370 $\mu$m (1/$e^{2}$ radius), respectively. So the resonant Rabi frequency of the pump beam is $2\pi\times 535 MHz$ \cite{CesiumData}. The optical power of the seed beam is 9.2 $\mu$W. After the FWM process, the optical powers of the amplified probe beam and the newly generated conjugate beam are 57.6 $\mu$W and 50.0 $\mu$W, respectively.

A Glan-Thompson (GT) polarizer with a  discrimination of $10^{5}$:1 is used after the vapor cell to filter out the pump beam. The amplified probe and the generated conjugate beams are directly sent into the two ports of a balanced photodetector (BPD) with a gain of $10^{5}$ V/A and a quantum efficiency of 98\%. A spectrum analyzer (SA) is used to record the output of the BPD. 

A flip mirror (FM) is used after the GT polarizer for imaging the beam patterns of the seed, probe, and conjugate beams on a charged-coupled device (CCD) camera. The beam pattern of the seed beam is imaged by blocking the pump beam before the vapor cell, while the beam patterns of the probe beam and conjugate beam are imaged at the same time when the pump beam is unblocked [upper panel of Fig. 1(c)]. The bright spot between the probe beam and conjugate beam is the leakage of the strong pump beam. Interference patterns of these three beams with plane wave beams are also taken by the camera by interfering each beam with a plane wave beam at the same frequency [lower panel of Fig. 1(c)]. The plane wave beams at two different frequencies (probe beam frequency and conjugate beam frequency, respectively) are generated by another FWM process in the same vapor cell, in which a Gaussian beam at probe beam frequency is seeded. Our measured results agree well with the ideal beam patterns and interference patterns for the case in which the seed beam carries topological charge $l=-1$, as shown in Fig. 1(b).

\begin{figure}[t]
\centering
\fbox{\includegraphics[width=\linewidth]{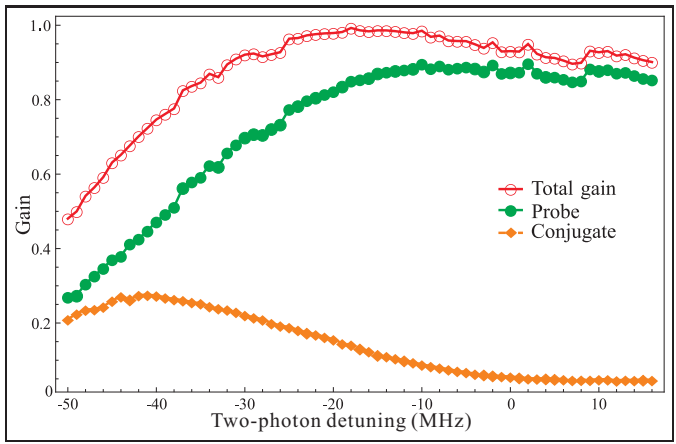}}
\caption{Gains of the probe beam and conjugate beam and their sum as a function of two-photon detuning $\delta$. Experimental parameters are as follows: $\Delta = 1.6$ GHz, $P_{Pump}$ = 740 mW, and $T$ = 80 $^{\circ}$C.} 
\end{figure}

Fig. 2 shows the noise power levels of individual probe beam (trace $B$), conjugate beam (trace $C$), and their relative intensity difference (trace $D$) measured at the analysis frequency of 1.2 MHz. 
All of these three traces are normalized to the corresponding SNL (trace $A$). As we can see, trace $B$ and trace $C$ are $8.1\pm0.1$ dB and $8.5\pm0.1$ dB above the corresponding SNL, respectively, because the noise of the probe beam and conjugate beam is amplified in the FWM process. Trace $D$ is $4.9\pm0.1$ dB below the SNL. It shows that quantum correlated twin beams carrying OAMs are generated.

Next, we investigate the classical behavior of the FWM process operated in a nonamplifying regime \cite{QuentinPRA}. The main difference from the high gain regime is the choice of vapor cell temperature $T$ = 80 $^{\circ}$C, so that the atomic density is around one order lower than that at $T$ = 112 $^{\circ}$C. Besides, the pump power is improved to 740 mW. Fig. 3 shows the gains of the probe beam and conjugate beam as well as the total gain of the twin beams as a function of two-photon detuning $\delta$, which is realized by tuning the EOM driving frequency. As the two-photon detuning decreases from $+16$ MHz to $-50$ MHz, the gain of the probe beam first keeps almost constant at 0.85, and then starts to decrease drastically at $-15$ MHz until it reaches 0.27 at $-50$ MHz. On the other hand, the gain of the conjugate beam reaches maximum 0.27 at $-42$ MHz and then decreases slightly when the two-photon detuning decreases further. As a result, the total gain of the twin beams is equal to or slightly smaller than one (specifically $>0.95$) in the two-photon detuning range $-25$ MHz to $-5$ MHz, and its maximal value one is achieved at around $-20$ MHz. In other words, the sum of the twin beams' optical power is almost equal to or slightly smaller than the optical power of the seed beam within this region of parameter space. In this sense, the function of the FWM process in the nonamplifying regime is similar to an optical beam splitter.

\begin{figure}[t]
\centering
\fbox{\includegraphics[width=\linewidth]{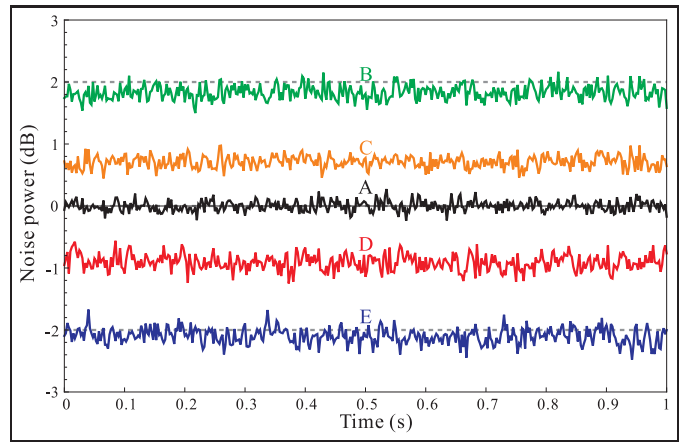}}
\caption{Intensity-difference squeezing of quantum correlated twin beams carrying OAMs generated from FWM process operated in nonamplifying regime. Traces $B$ and $C$ show normalized noise power of the probe beam and conjugate beam, respectively. Trace $D$ and $E$ show the normalized intensity-difference of the twin beams without and with attenuation on the probe beam, respectively. Trace $A$ at 0 dB shows the corresponding SNL of trace $B\sim E$. Experimental parameters are as follows: $\Delta = 1.6$ GHz, $\delta = -19$ MHz, $P_{Pump}$ = 740 mW, and $T$ = 80 $^{\circ}$C. Resolution bandwidth (RBW) of SA is 30 kHz and video bandwidth (VBW) of SA is 100 Hz.}
\end{figure}

Then we study the noise characteristics of the FWM process operated in the nonamplifying regime. The two-photon detuning is set at $-19$ MHz, and other experimental parameters are the same as in Fig. 3. Under these conditions, the gains of the probe beam and conjugate beam are 0.84 and 0.16, respectively. Trace $B$ and trace $C$ in Fig. 4 show the normalized noise power levels of individual probe beam and conjugate beam measured at the analysis frequency of 1.2 MHz, respectively. The probe beam and conjugate beam are $1.8\pm0.1$ dB and $0.7\pm0.1$ dB above the corresponding SNL, respectively. The noise power of the intensity difference of the twin beams carrying OAMs is $0.9\pm0.1$ below the SNL (trace $D$). So far, it is demonstrated that the probe beam and conjugate beam carrying OAMs generated from the FWM process operated in nonamplifying regime, in analogy to two output channels of an optical beam splitter, are quantum correlated. Then the FWM process operated in nonamplifying regime can be regarded as a quantum beam splitter for OAMs. The degree of quantum correlation can be increased to $2.1\pm0.1$ dB by adding 67\% loss on the probe beam (trace $E$). This can be understood by taking into account the strong power imbalance of the probe beam and conjugate beam. Extra loss on the probe beam makes these twin beams more balanced and thus improves the noise reduction on the degree of quantum correlation \cite{JasperseOE}.

\begin{figure}[t]
\centering
\fbox{\includegraphics[width=\linewidth]{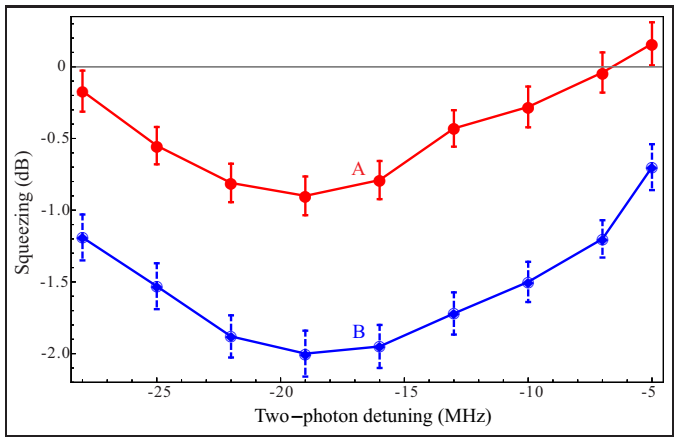}}
\caption{Degree of intensity-difference squeezing from the FWM process operated in nonamplifying regime as a function of two-photon detuning. Traces $A$ and $B$ show the normalized intensity-difference noise levels of the twin beams without and with attenuation on the probe beam, respectively.}
\end{figure}

Fig. 5 shows the dependence of quantum correlation on two-photon detuning in the low-gain regime. Traces $A$ and $B$ show the normalized intensity-difference noise levels of the twin beams without and with attenuation on the probe beam, respectively. It can be seen that quantum correlation exists within a wide two-photon detuning range from -28 MHz to -7 MHz. The gains of the probe beam and conjugate beam shown in Fig. 3 can be in analogy to effective transmission and reflection coefficients of an optical beam splitter. In this sense, the effective transmission and reflection coefficients of our quantum beam splitter can be manipulated within a certain range by tuning the two-photon detuning of the FWM process.

In conclusion, we experimentally demonstrated the generation of quantum correlated twin beams which carry nonzero OAMs based on the double-$\Lambda$ energy level FWM process in hot cesium vapor. The amplified probe beam and the newly generated conjugate beam carry opposite topological charges due to the conservation of angular momentum in the FWM process. Furthermore, we also investigated the case in which the total power of the twin beams carrying OAMs is equal to the power of the seed coherent beam, i.e., the FWM process is operated in a nonamplifying regime. Quantum correlation is also observed between the $l=-1$ probe beam and $l=+1$ conjugate beam in this nonamplifying regime. In this sense, the FWM process operated in nonamplifying regime works as a quantum beam splitter of OAMs. Our work paves the way to using FWM process as quantum beam splitter of images based on its multi-spatial-mode advantage. The FWM process operated in pulsed regime has been realized in hot rubidium vapor \cite{PulsedFWM}, which paves the way to its applications in quantum communication. Generating quantum correlated twin beams carrying higher order OAMs operated in pulsed regime is also attractive, as higher order OAM has the potential to improve the capacity of quantum communication.

This research was supported by the National Natural Science Foundation of China (Grants No. 61601270, and No. 11834010), National Key R\&D Program of China (Grants No.
2016YFA0301402), Fund for Shanxi ``1331" Project Key Subjects Construction.

\end{document}